\documentclass[final,authoryear,times]{elsarticle}
\usepackage{graphicx}
\usepackage{amssymb}
\journal{Transportation Research Part C}

\usepackage{amsmath}
\usepackage{units}

\begin{document}

\begin{frontmatter}

\title{Calibration and validation of models describing the spatiotemporal evolution of congested traffic patterns}

\author[TUD]{Martin Treiber\corref{cor1}}\ead{treiber@vwi.tu-dresden.de}
\author[TUD]{Arne Kesting}\ead{kesting@vwi.tu-dresden.de}

\address[TUD]{Technische Universit\"at Dresden, Institute for Transport \& Economics,\\
W\"urzburger Str. 35, 01062 Dresden, Germany}
\cortext[cor1]{Corresponding author.}

\begin{abstract}
We propose a quantitative approach for calibrating and validating key features of traffic instabilities based on speed time series obtained from aggregated data of a series of neighboring stationary detectors. We apply the proposed criteria to historic traffic databases of several freeways in Germany containing about~400 occurrences of congestions thereby providing a reference for model calibration and quality assessment with respect to the  spatiotemporal dynamics. First tests with microscopic and macroscopic models
indicate that the criteria are both robust and discriminative, i.e., clearly distinguishes between models of higher and lower predictive power. 
\end{abstract}

\begin{keyword}
Spatiotemporal dynamics \sep traffic instabilities
\sep calibration \sep validation \sep traffic flow models \sep stop-and-go traffic

\end{keyword}

\end{frontmatter}


\section{Introduction}\label{sec:intro}
%
Traffic flow models displaying instabilities and stop-and-go waves
have been proposed for more than fifty years~\citep{Reu50a,Pipes} 
(see also the reviews by~\cite{Helb-opus,Hoogendoorn-review}). 
Observations of instabilities date back several
decades as well~\citep{Treiterer}. Nevertheless, a systematic concept for
quantitatively calibrating and validating model predictions of instabilities against
observed data seems still to be missing. A possible reason for this are
intricacies in the data interpretation and wrong expectations about
what can be achieved and what not.

In principle, traffic instabilities and oscillations in congested
traffic can be captured by several types and
representations of data: flow-density points
and speed time series of aggregated stationary detector data (SDD), 
passage times and individual speeds from single-vehicle SDD,
floating-car data captured by the vehicles themselves, and
trajectory data derived from images of cameras at high viewpoints. 

\begin{figure}

\centering
\includegraphics[width=0.8\textwidth]{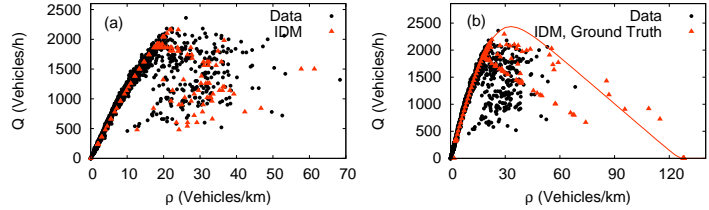}
\caption{\label{fig:fund}(a) Comparison of empirical and simulated
flow-density data; (b) simulated ground truth for the density with the modeled fundamental diagram (solid
curve).}
\end{figure}

At first sight, flow-density data seem to be useful for representing
traffic instabilities since oscillations are non-stationary phenomena in which traffic is 
clearly out of equilibrium. This should lead to a wide scattering
in the flow-density data of congested traffic~\citep{Kerner-Rehb96-2} as shown in
Fig.~\ref{fig:fund} regardless of the theoretical question whether
traffic flow has a definite equilibrium relation (``fundamental
diagram''), or not~\citep{Katsu03,Kerner-book,Treiber-ThreePhasesTRB}. 

In fact, simulations with the Intelligent Driver
Model~\citep{Opus}, as well as with a wide variety of other models, show 
wide scattering even for identical drivers and vehicles (Fig.~\ref{fig:fund}). 
Nevertheless, flow-density data cannot be
used for a quantitative calibration since other effects such as inter-driver 
variability (heterogeneous traffic)~\citep{Banks-99,Ossen-interDriver06,GKT-scatter}, intra-driver
variability (drivers change their behavior over time)~\citep{IDMM}, and many
effects related to lane changes and platoons~\citep{Katsu03,VDT} may lead to wide scattering as
well. 

Moreover, since the density is not directly measurable by
stationary detectors, it must be estimated, typically by making use of
 the hydrodynamic relation ``flow divided by the average
velocity''. The available temporal average, however, systematically
underestimates the true spatial average relevant for this relation which
results in an extreme bias in the flow-density data of congested
traffic~\citep{Leutzbach,Helb-opus}. Figure~\ref{fig:fund} illustrates this by means of
simulations with the Intelligent Driver
Model by~\cite{Opus}). This model has been calibrated such that the
aggregated virtual detector shows a similar scattering as the empirical
data (Fig.~\ref{fig:fund}(a)). Plotting the same data points using the
true spatial density (which, of course, is available in the
simulation) reveals the systematic errors (Fig.~\ref{fig:fund}(b)).

Alternatively, detailed information can be extracted from single-vehicle data,
floating-car data, or trajectory data. 
However, since oscillations are collective phenomena,
the high level of detail is not necessary. Moreover, to date, the
availability of these data types is still limited, or the data are not
suitable for our purposes. Specifically, the NGSIM trajectory data~\citep{NGSIM}
widely used as a testbed for these types of data
cover road sections of typically~\unit[2]{km}  in length while
stop-and-go waves develop on typical scales of \unit[5-10]{km}.

\begin{figure}
\centering
\includegraphics[width=0.7\textwidth]{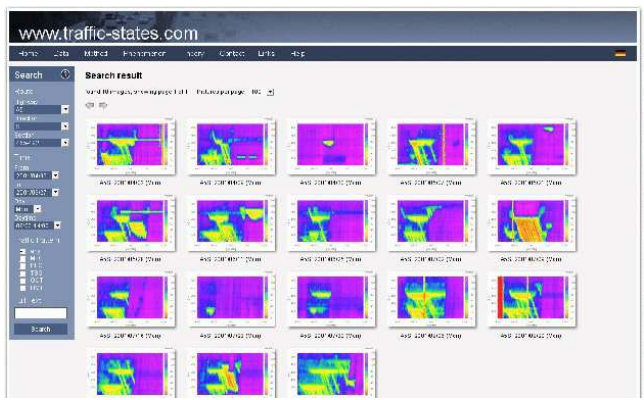}
\caption{\label{fig:traffic-states.com}{\small Screenshot of a publicly
available image database of congestions on the German freeway A5. The database 
 can be searched by criteria such as direction of travel, space
interval, time interval,  weekday, cause of the congestion, 
or the type of traffic pattern. In the example, the database has been
filtered for jams occurring on Mondays in the southern driving direction. The
images have been produced using a dedicated reconstruction technique~\citep{ASM-CACAIE}.}
}
\end{figure}

In contrast, speed time series obtained from aggregated stationary
detector data are widely
available (see Fig.~\ref{fig:traffic-states.com}) 
and thus suitable for systematic calibration and
benchmarking purposes. Moreover, since temporal averages are the
appropriate mean when analyzing time series, no systematic errors
arise. Using this type of data, the wavelength and propagation velocities
of oscillations on freeways has been investigated by~\cite{Lindgren-PhD-A5,BertiniLeal-M4,Ahn-Cassidy-OscillationsISTTT,Martin-empStates,Wilson-M42} and compared for several countries by~\cite{Zielke-intlComparison}. 
However, the proposed schemes are not suitable for a systematic automated
analysis. Moreover, to our knowledge, there seem to be no published
work quantitatively determining the growth rate of the oscillations.

In this paper, we propose a systematic approach for calibrating and validating key 
features of traffic instabilities based on speed time series obtained
from aggregated stationary detector data. This includes the
propagation velocity, the wavelength, and the growth rate of the
oscillations. We give first results
based on a historic congestion database of the German freeway A5,
and show that the method has a high discriminative power in
assessing the model quality.

The proposed criteria are described in Sec.~\ref{sec:methods},
and applied to observed data in Sec.~\ref{sec:A5}. A first application to traffic flow models is given
in Sec.~\ref{sec:models} before Sec.~\ref{sec:conclusion}
concludes with a discussion and an outlook.

\section{\label{sec:methods}The Proposed Quality Measures}
%
The formulation of the criteria is based on qualitative empirical
findings (``stylized facts'')~\citep{Treiber-ThreePhasesTRB} that are  
persistently observed on various freeways all over the world, e.g.,
the USA, Great Britain, the Netherlands, and Germany~\citep{Lindgren-PhD-A5,BertiniLeal-M4,Ahn-Cassidy-OscillationsISTTT,Martin-empStates,Zielke-intlComparison,Helb-Phases-EPJB-09,Wilson-Pattern2008}. For our purposes, the relevant stylized facts  are the following:

\begin{enumerate}
\item \emph{Congestion patterns 
are typically caused by bottlenecks} which may be caused, e.g., by
intersections, accidents, or gradients. 
Analyzing about 400 congestion patterns on the German
freeways A5-North and A5-South 
(Fig.~\ref{fig:traffic-states.com}) did not bring conclusive evidence
of a single breakdown without a bottleneck~\citep{Martin-empStates}.

\item \emph{The congestion pattern can be localized or spatially
extended}. Localized patterns, as well as the downstream boundary of
extended congestions,  are either pinned at the bottleneck,
or moving upstream at a
characteristic velocity $c_{\rm cong}$. Only extended patterns
(Fig.~\ref{fig:timeseriesOCT}) and moving localized patterns
 (isolated stop-and-go waves, Fig.~\ref{fig:timeseriesMLC}) 
 are relevant for our investigation. 

\item \emph{Most extended traffic
 patterns exhibit internal oscillations} propagating upstream approximately at 
the same characteristic velocity 
$c\approx c_{\rm cong}=\text{const}$~\citep{Dag99a,Mauch-Cassidy}.
 In spatiotemporal representations such as that of Fig.~\ref{fig:timeseriesMLC} this
corresponds to nearly parallel structures. For free traffic, the propagation
velocity $c$ is somewhat lower than the average vehicle speed $\overline{V}$.

\item \emph{The amplitude of the oscillations increases while
propagating upstream}~\citep{BertiniLeal-M4,Ahn-Cassidy-OscillationsISTTT}. 
We quantify this by defining a temporal growth
rate $\sigma$ characterizing the growth process before saturation.

\item \emph{The average wavelength $L$ of the (saturated) waves increases
with decreasing bottleneck strength}~\citep{Opus,Treiber-ThreePhasesTRB}.
\end{enumerate}
In formulating the calibration criteria, 
We use the first two facts to filter the input data for applicable
spatiotemporal regions (Sec.~\ref{sec:filter}) 
while the quantities $c$, $\sigma$, and $L$
quantifying  the remaining stylized facts constitute the criteria itself
(Sections~\ref{sec:c}-\ref{sec:L}). In general, $c$, $\sigma$, and $L$
depend on the bottleneck strength which, therefore, needs to be
quantitatively defined as well 
(Sec.~\ref{sec:bottlStrength}).

\subsection{Data Description}

We assume that aggregated data for the flow $Q$ and speed $V$
(arithmetic average) are available from 
several detector cross sections at locations $i$. The distance between
two adjacent detectors should not exceed about \unit[2.5]{km} which
generally is satisfied on relevant freeway sections (i.e., sections
where traffic breakdowns are observed regularly).
Usually, the flow and speed data $(V_{ilt}, Q_{ilt})$ of cross section $i$ 
are available for each lane (index $l$) after each time interval 
(index $t$). With few exceptions, congested traffic is
``synchronized'' between lanes, i.e., the average speed at a given
location and time is not significantly different across the lanes. We
therefore use the following 
weighted averages as input for our calibration criteria:
\begin{equation}
\label{laneAverage}
V_{it}=\frac{1}{Q_{it}}\sum_{l=1}^{n_i}Q_{ilt}V_{ilt}
\end{equation}
where $n_i$ is the number of mainroad lanes at location $i$. The 
total flow is given by
\begin{equation}
\label{Qtot}
Q_{it}=\sum_{l=1}^{n_i}Q_{ilt}.
\end{equation}

\subsection{\label{sec:filter}Filtering Spatiotemporal Regions}
\begin{figure}
\centering
\centering
\includegraphics[width=0.9\textwidth]{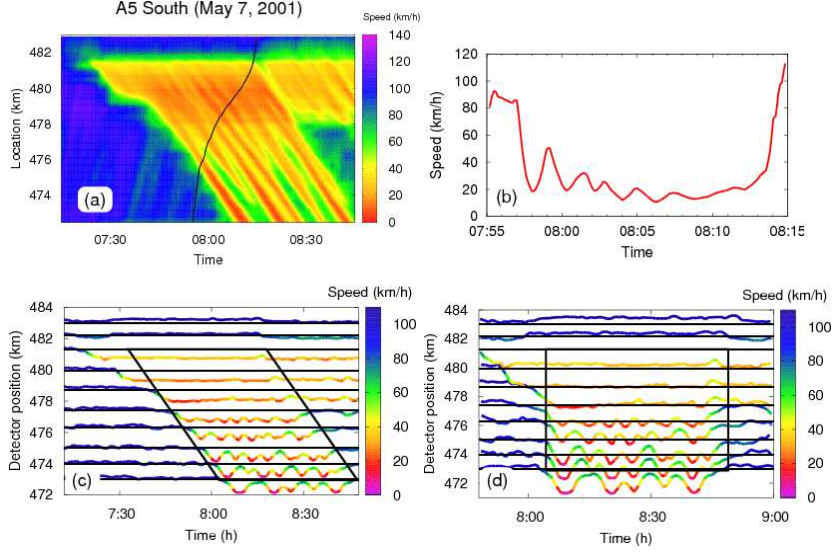} 
\caption{\label{fig:timeseriesOCT}Example of stationary detector data 
displaying the development of stop-and-go waves from stationary
congested traffic near the bottleneck. 
(a)~Reconstruction of the spatiotemporal evolution of local vehicle
speeds using a dedicated reconstruction
technique~\citep{ASM-CACAIE}. Also shown is a reconstructed vehicle
trajectory (black curve). 
(b)~Speed-time series of the reconstructed
trajectory. 
(c)~Speed-time series according to Eq.~\eqref{laneAverage}.
The locations of the respective detectors are
given by the black baselines, and the speed (reference value~\unit[90]{km/h}) 
is plotted relative to the baseline. The  black parallelogram encloses 
the relevant spatiotemporal range defined in Sec.~\ref{sec:filter}.
(d)~Representation of the same data with a
skewed time axis $t'=t-(x-x_1)/c_\text{cong}$ with $x_1=\unit[473]{km}$ and
$c_\text{cong}=\unit[-16]{km/h}$.
}
\end{figure}

\begin{figure}
\centering
\includegraphics[width=0.9\textwidth]{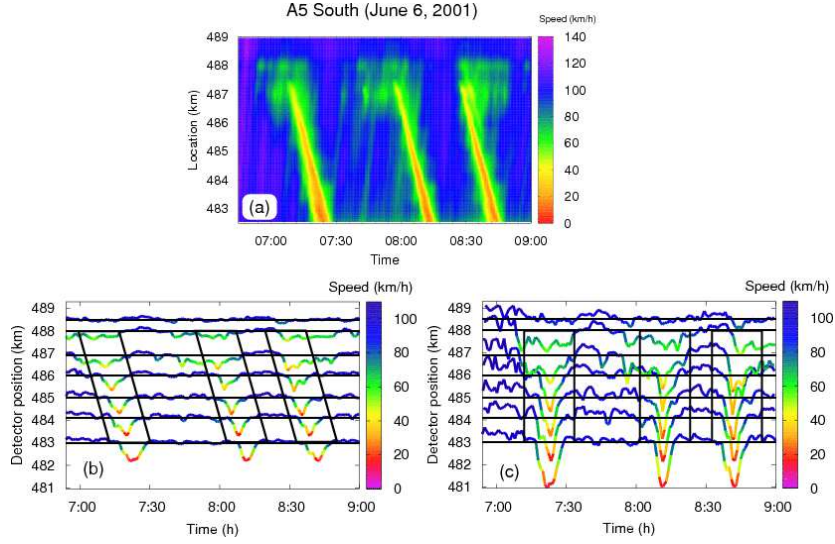} 
\caption{\label{fig:timeseriesMLC}Stationary detector data displaying
the evolution of isolated stop-and-go waves caused by a weak
bottleneck located at about \unit[487.5]{km}. (a)~The spatiotemporal speed field, 
(b)~the time series, and (c)~the time series with skewed time axis 
are derived from the data as in Fig.~\ref{fig:timeseriesOCT}. The parameters of the skewed time
transformation $t'=t-(x-x_1)/c_\text{cong}$
are given by $x_1=\unit[483]{km}$ and $c_\text{cong}=\unit[-16]{km/h}$.}
\end{figure}

In order to determine the dynamic quantities $c$, $\sigma$, and $L$ from the
input data $\{V_{it}\}$ it is necessary to restrict the spatiotemporal
range to regions of actual congestions. From the stylized facts above,
we use the information that the downstream boundary of congestions is
either stationary at a bottleneck, or moving with a constant velocity
$c_\text{cong}$. Furthermore, internal structures move at nearly the same
velocity. This leads to defining parallelogram-shaped regions for both
isolated stop-and-go waves, and extended congestions (cf.\ Figs.~\ref{fig:timeseriesMLC} 
and~\ref{fig:timeseriesOCT}). The parallelogram
is defined as follows:
\begin{itemize}
\item Two edges correspond to detector locations. The downstream
edge is defined by the 
detector that is nearest to the bottleneck (in upstream direction).
The upstream edge is at the detector location where the waves begin
to saturate. This should include $n\ge 3$ detector cross sections. In
the following, we will count the cross sections in the direction of
the flow, i.e., the upstream edge is located at $x_1$, and the
downstream edge at $x_n$.
\item The other two edges are angled such that, in the spatiotemporal
picture, the waves propagate parallel to these edges.
\item The vertices $(x_1,t_1^\text{beg})$ and $(x_1,t_1^\text{ end})$ connected
by the upstream edge are defined such that the parallelogram has a
maximum area subject to the condition that there is no free traffic inside:
\begin{eqnarray}
t_1^\text{beg} &=& \max\limits_i \left(\hat{t}_i-\frac{x_i-x_1}{c_\text{cong}}\right),\\
t_1^\text{ end} &=& \min\limits_i
\left(\tilde{t}_i-\frac{x_i-x_1}{c_\text{cong}}\right) := t_1^\text{beg}+T\,.
\end{eqnarray}
Here, $\hat{t}_i$ is the time where the aggregated speed time series $V_{it}$ of
cross section $i$ drops below a critical speed $v_c$,
and  $\tilde{t}_i$ is the time where it rises above $v_c$. To reduce the
occurrence of
false signals, the time series can be smoothed over time scales of a
few minutes just for this step. A value $v_c=\unit[70]{km/h}$
turned out to give good results for freeways.
\end{itemize}

\subsection{\label{sec:bottlStrength}Quantifying the Bottleneck Strength}

For extended congestions, 
the quantities $c$, $\sigma$, and $L$ depend on the bottleneck
strength. Therefore, this exogenous variable has to be characterized by the
measured quantities as well. Theoretically, the bottleneck strength is
defined by a local drop of the capacity that is available for the  mainroad
traffic~\citep{Martin-empStates}. However, this quantity is
problematic to measure, particularly, if the bottleneck includes
merging/diverging traffic (e.g., junctions, intersections, or lane
closings). Therefore, we use as proxy the average vehicle speed at
the detector nearest to the bottleneck,
\begin{equation}
\label{Vbottl}
\overline{V}=\frac{1}{T+1}
\sum\limits_{t= t_1^\text{beg}+\Delta t}^{t_1^\text{ end}+\Delta t}V_{nt}, 
\quad
\Delta t=\frac{x_n-x_1}{c_\text{cong}}.
\end{equation}
At this location, the oscillations inside extended jams are minimal
(Stylized Fact~4) making $\overline{V}$ a well-defined quantity. 
In order to plot the characteristic quantities of isolated stop-and-go
waves into the same diagram, such as Fig.~\ref{fig:sysProperties}, we define
$\overline{V}$ as the speed 
\emph{outside} of the wave, for this case.

\subsection{\label{sec:c}Propagation Velocity}

The {propagation velocity} $c$ can be
calculated by
maximizing the sum of cross-correlation functions  of speed time
series of  detector pairs  $\{i,j\}$ in the congested region (at positions $x_i$ and $x_j$,
respectively), with
respect to the velocity $c$~\citep{Coifman-Wang-CrossCorr-ISTTT,Zielke-intlComparison}:
\begin{equation}
\label{corr}
c=\text{arg}\, \max_{c'}
\sum\limits_i \sum \limits_{j>i} \text{Corr} \left[V_i(t), V_j \left(t+\frac{x_i-x_j}{c'}\right)\right],
\end{equation}
where the continuous-in-time function $V_i(t)$ is defined by a
piecewise linear interpolation of the detector time series
$V_{it}$. Stylized Fact~3 suggests that the values of $c$ lie in a
very small range which is confirmed by Fig.~\ref{fig:sysProperties}(d).
 This makes it possible to define the spatiotemporal
region using the a-priori estimate $c_\text{cong}$ without danger of
circular reasoning. When selecting spatiotemporal regions of free
traffic, Eq.~\eqref{corr} can be used to determine the propagation
velocity of perturbations in free traffic (data points at the upper right corner of
Fig.~\ref{fig:sysProperties}(d)). 

\subsection{\label{sec:sigma}Growth rate}

The average spatial growth rate $\tilde{\sigma}$
of perturbations is defined in terms of the slope of the linear regression of
the logarithm of the amplitude,
\begin{equation}
\label{spatialgrowth}
\tilde{\sigma}=\frac{\sum_i x_i \ln |A_i|-n\overline{x} \, \overline{\ln |A_i|}} 
 {\sum_i x_i^2-n\overline{x}^2},
\end{equation}
where the amplitude $A_i$ is approximated by the standard deviation of
the speed time series $V_{it}$ inside the parallelogram.
Of course, $\tilde{\sigma}$ is only a rough approximation of the true
growth rate. Particularly, $\tilde{\sigma}$ is influenced by 
non-collective fluctuations caused by random noise and driver-vehicle
heterogeneity as well as by saturation effects, particularly, if the
spatiotemporal region is not defined carefully. Both effects
systematically decrease the estimated growth rate but do not
invalidate the overall picture. However, it is
essential to calculate the simulated quantities using exactly the same
measuring and estimation procedures.

Once the spatial growth rate is known, the temporal growth rate is
given by
\begin{equation}
\label{sigmaEmp}
\sigma=c \tilde{\sigma}.
\end{equation}
Notice that, in general, $\tilde{\sigma}$ is negative, and $\sigma$ positive.
%

\subsection{\label{sec:L}Wavelength}

The average period  $\tau$ of the oscillations of extended congestions
is given by the
position of the first nontrivial  peak of the autocorrelation 
functions of detector time
series inside the parallelogram. In some cases,  waves merge during
their propagation such that the period and the wavelength increases with the distance
from the bottleneck. We therefore define $\tau$ at a certain
development stage, namely at the limit of saturation:
\begin{equation}
\label{tauEmp}
\tau=\text{arg}\, \max_{\tau'>0}\, \text{Corr} \left[V_1(t), V_1(t+\tau') \right].
\end{equation}
In analogy to Eq.~\eqref{sigmaEmp}, the associated wavelength is given by
\begin{equation}
\label{LEmp}
L=|c| \tau.
\end{equation}

\section{\label{sec:A5}Application to Freeway Data}

The characteristic quantities defined in the previous section have
been calculated for about~30 instances of traffic congestions observed on a
section of  the
German freeway A5-South near Frankfurt for April and May~2001.
 Images of the corresponding 
spatiotemporal speed fields can be viewed at {\tt
www.traffic-states.com} (cf.\ Fig.~\ref{fig:traffic-states.com} for a
screenshot). Not all of the more than 60~jams observed in the
southern direction in this time period 
are suited for the analysis: Localized jams must be
moving (filter criterion ``MLC''), and extended jams must exhibit
distinct oscillating structures (filter criteria ``OCT'' or 
``TSG''). Additionally, they must be sufficiently extended such that
the parallelogram includes at least three cross sections, and each
cross section records at least three  oscillations. While oscillating
structures are observed for most of the cases except 
for very strong or weak bottlenecks (filter criteria ``HCT'' and
``HST'', respectively), nearly \unit[50]{\%} of the remaining
candidates do not satisfy the minimum size criteria.

\begin{figure}
\centering
\includegraphics[width=1.0\textwidth]{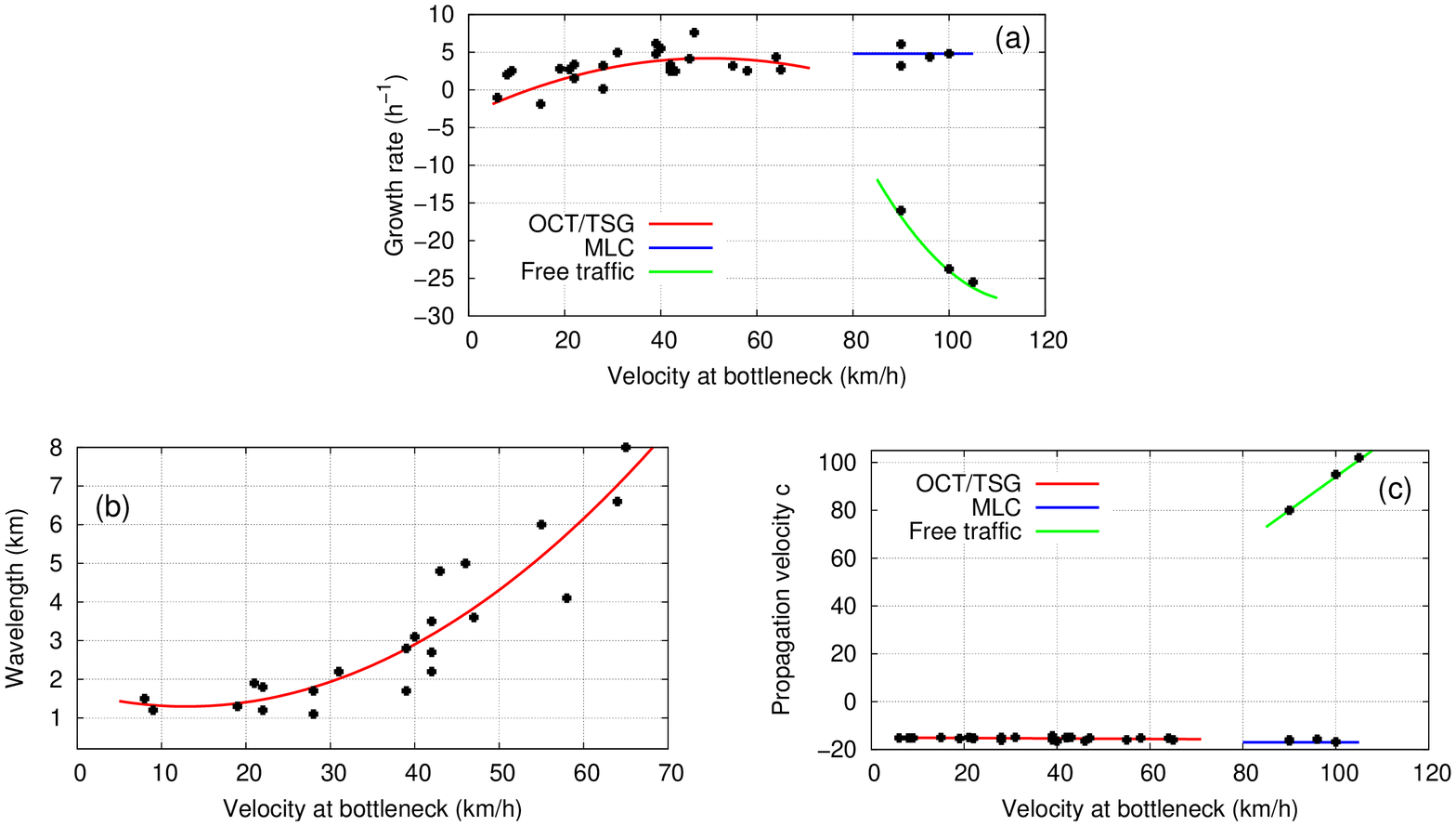}
\caption{\label{fig:sysProperties}Characteristic properties of the
propagation of perturbations in congested traffic. (a)~Growth rate
$\sigma$, (b)~wavelength $L$, and (c)~propagation velocity $c$, as a
function of the average speed $\overline{V}$ characterizing the bottleneck strength.
}
\end{figure}

Figure~\ref{fig:sysProperties} displays the main result which can be
used not only for calibration purposes, but also for understanding the
traffic dynamics in general. Part (a) of this figure shows the growth
rate of the moving structures inside extended congestions (points for
$\overline{V}$ below \unit[70]{km/h}), the growth rate of isolated
stop-and-go waves (MLC, top right corner), and, for comparison, the growth
rate of some moving structures in free traffic. For congested traffic,
the growth rate is generally positive, apart for very low average
velocities $\overline{V}$ corresponding to extended congestions behind 
strong bottlenecks (usually
accident-caused lane closures, selected in the database 
by the  criterium ``accident''). This positive growth rate can be
directly experienced by the driver (or alternatively obtained from
floa\-ting-car data) as shown in Fig.~\ref{fig:timeseriesOCT}(b): When
approaching the bottleneck , the experienced velocity oscillations 
apparently decrease in amplitude with the growth rate
\begin{equation}
\label{sigmaFC}
\sigma_\text{FC}=\overline{V}_\text{FC}\tilde{\sigma}.
\end{equation}
They have the apparent period 
\begin{equation}
\label{TFC}
\tau_\text{FC}=\tau\left(\frac{|c|}{|c|+\overline{V}_\text{FC}}\right)
\end{equation}
where $\overline{V}_\text{FC}$ is the average speed of the floating car.

Figure~\ref{fig:sysProperties}(b) shows that the wavelength decreases
with increasing bottleneck strength, but always remains above~\unit[1]{km}. 
Finally, Fig~\ref{fig:sysProperties}(c) confirms that
the propagation velocity is nearly constant (in the range
$\unit[-17]{km/h}$ to $\unit[-16]{km/h}$ for extended jams and about 
$\unit[-18]{km/h}$ for isolated moving waves).

\begin{figure}
\centering
\includegraphics[width=0.5\textwidth]{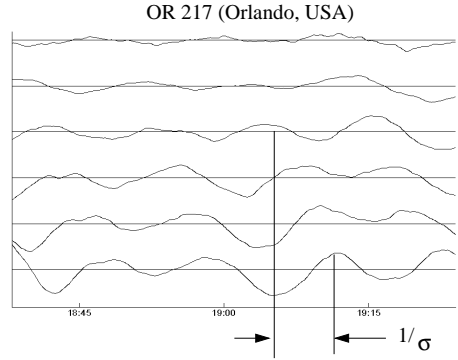}
\caption{\label{fig:benZielke}Speed-time series of several
detectors on the US~American freeway OR~217~\citep{Zielke-intlComparison}. 
Shown are deviations (oscillating curves) from the average speed
(horizontal baselines). 
The detector spacings are proportional to the spacings of the
baselines. Also shown is the time period $1/\sigma$ in which the
amplitudes of each wave increase by a factor of~$e$.}
\end{figure}

The analysis  quantitatively confirms the
stylized facts presented in Sec.~\ref{sec:methods}.
Particularly, there is no empirical evidence
against the existence of linear instabilities in congested traffic.
This is also confirmed by data from other freeways in several
countries, although there are quantitative deviations (this, exactly,
is the basis for calibrating the models). As an example,
Fig.~\ref{fig:benZielke} taken from~\cite{Zielke-intlComparison}
shows speed time series which are suitable
for the present analysis. The growth rate is comparable to the A5 data
while the propagation velocity $c\approx \unit[-20]{km/h}$ 
has a somewhat higher absolute value~\citep{Zielke-intlComparison} .

\section{\label{sec:models}A First Test of Traffic Flow Models}

In this section, we will show how the results can be applied to
calibrating the dynamic aspects of traffic flow models by 
testing the Intelligent-Driver Model (IDM) \citep{Opus} and the Human
Driver Model (HDM) \citep{HDM} on the extended congestion shown in 
Fig.~\ref{fig:timeseriesOCT}.
Both models are time-continuous 
car-following models, but the calibration procedures can be applied as
well to
discrete-in-time models, cellular automata and to macroscopic models.
 
\begin{figure}
\centering
\includegraphics[width=0.9\textwidth]{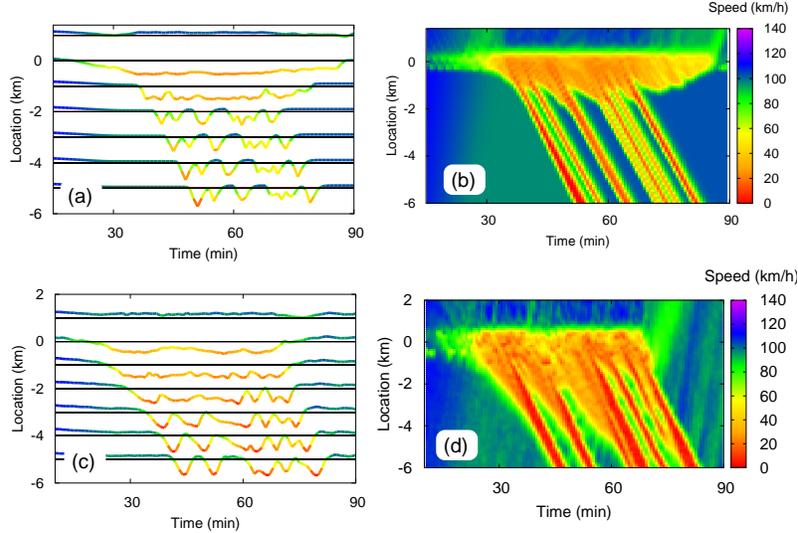}
\caption{\label{fig:simOCT}
Examples of simulated virtual detector data based on simulations with
the Intelligent Driver Model for a permanent bottleneck (plots~a and~b), and the Human Driver
Model with a temporary bottleneck (c, d). The simulated data
aggregation (time intervals of \unit[1]{min}), and the transformations
to display the data are the same as for the empirical data in
Fig.~\ref{fig:timeseriesOCT}.}
\end{figure}

In the simulations, the bottleneck is represented by an on-ramp. The
on-ramp flow, i.e., the bottleneck strength, 
is chosen such that the average velocity $\overline{V}$ near the
bottleneck, Eq.~\ref{Vbottl}, is consistent with the data. At
$t=\unit[70]{min}$, the bottleneck is reduced resulting in a moving 
downstream front of the congestion at later times (this has no
influence on the calibration results).
Virtual detectors are positioned every kilometer, and vehicle passage
times and speeds are
aggregated in one-minute intervals, in analogy to the data sampling at
the real detectors. The result is displayed in Fig.~\ref{fig:simOCT}.

The criteria for the actual calibration and assessment are the
propagation velocity $c$, wavelength $L$ (or period $\tau$), and growth rate
$\sigma$. They are calculated from the virtual detector data with
exactly the same scheme that has been used 
for the real data. Since the speed $\overline{V}$, i.e., the bottleneck strength, is
fixed, there are no degrees of freedom by manipulating the
bottleneck. A comparison with the empirical values
$\sigma_\text{obs}=\unit[5]{h^{-1}}$,
$c_\text{obs}=\unit[-17]{km/h}$, and $\tau_\text{obs}=\unit[10]{min}$
(cf. Fig.~\ref{fig:timeseriesOCT}) gives a first
impression of the discriminative power of the proposed calibration and
testing method: The empirical propagation velocity $c$ agrees with both
models. However, for the IDM simulation, the period $\tau_\text{IDM}=\unit[6]{min}$
of the waves is too
short, and the growth rate $\sigma_\text{IDM} \approx \unit[10]{h^{-1}}$
(calculated using the detectors at 0, 1, and \unit[2]{km}) is too
high. It turns out that this cannot be improved significantly by
changing the model parameters, i.e., calibrating the model. For the
HDM simulation, the period $\tau_\text{HDM}=\unit[8]{min}$ deviates
less, and the growth rate $\sigma_\text{HDM} \approx
\unit[5]{h^{-1}}$ is consistent with the data. Since, in contrast to
the IDM,  the HDM includes explicit reaction times and
multi-anticipation (the drivers take into account multiple leaders),
this may mean that both aspects are essential for a quantitative
modelling. However, it is too early to draw definitive conclusions. 

\begin{figure}
\centering
\includegraphics[width=0.6\textwidth]{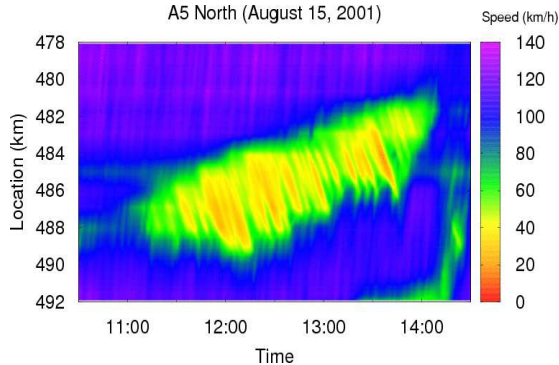}  

\caption{\label{fig:moving}A moving bottleneck filtered from the
image database \texttt{www.traffic-states.com}: Validating a model
with data it most probably has never seen before.}
\end{figure}

\section{\label{sec:conclusion}Discussion and Outlook}

In conclusion, we propose three aspects of traffic instabilities that
are accessible to a systematic and quantitative test of model
prediction: the wavelength, the propagation velocity, and the growth
rate.  Since all properties are functions of
the average velocity indicating the bottleneck strength, there should be enough
input for a discriminative testing of models. 
All quantities are defined in terms of automated schemes enabling
the systematic investigation of large-scale data bases. As a first
step, it is planned to investigate the complete database displayed at
{\tt www.traffic-states.com} containing more than 400 instances of
congestions. 

The database contains also some unusual patterns to which 
the scheme of Sec.~\ref{sec:methods} can be easily generalized and which
may be used to validate the predictive power of models calibrated to
the more conventional patterns discussed above. For example, there are
some searchable instances of 
``moving bottlenecks'' (Fig.~\ref{fig:moving}), possibly caused by
very slow special transports or obstructions due to ``moving road
works'' (such as mowing the median
lawn). 

We notice that all proposed quantities for calibrating and testing 
relate to existing
congestions. Due to the stochastic nature of the perturbations
eventually leading to a breakdown, predictions for an actual breakdown
are necessarily restricted to a probabilistic
level~\citep{ElefteriadouProbBreakdown-1995,Brilon-stochCapa-2005}. 

It must be emphasized that
calibrations based on microscopic quantities such as the difference between
measured and simulated gaps are strongly influenced by unpredictable intra- and
inter-driver variations~\citep{Ossen-interDriver06,Kesting-Calibration-TRR08}
resulting in little discriminative power~\citep{Brockfeld-benchmark04,Ranjitkar-bench04,Ossen-benchmark05,Punzo-bench05}.
However, when calibrating against integrated quantities such as the
traveling time~\citep{Brockfeld-benchmark03}, important aspects such as accelerations or the
properties of oscillations are averaged away resulting in reduced
discriminative power as well. In contrast,  the schemes proposed here
probe the traffic dynamics on the intermediate scale of individual oscillations which is
of the order of one kilometer and a few minutes. Since we probe on
collective phenomena, inter- and intra-driver
variabilities should only play a minor role. However, the scale is 
sufficiently microscopic to retain accelerations. We therefore expect 
that the proposed schemes can be applied to a better and more
discriminative benchmarking of
microscopic and macroscopic models with respect to traffic
instabilities. This is the topic of ongoing work.

\subsection*{Acknowledgements}
The authors would like to thank the {\it Hessisches Landesamt f\"ur
Stra{\ss}en- und Verkehrswesen} for providing the freeway data.



\end{document}